\documentclass[conference]{IEEEtran}
\usepackage[switch]{lineno} 

\usepackage{cite}
\usepackage{amsmath,amssymb,amsfonts}
\usepackage{algorithmic}
\usepackage{algorithm}
\usepackage{graphicx}
\usepackage{textcomp}
\usepackage{xcolor}
\usepackage{url}
\usepackage{booktabs}
\usepackage{multirow}
\usepackage[keeplastbox]{flushend}

\usepackage{color}
\usepackage{framed}
\usepackage{comment}
\usepackage{hyperref}

\newcounter{t0d0_counter}
\newcommand{\nofixme}[1]{
}
\newcommand{\fixme}[1]{
 \stepcounter{t0d0_counter}
 \definecolor{shadecolor}{rgb}{1,1,0} 
 \begin{shaded}
 T0D0 \arabic{t0d0_counter}: #1
 \end{shaded}
}

\newcommand{\dumptenthirty}{\texttt{dump1030}}

\IEEEoverridecommandlockouts

\begin{document}

\title{\dumptenthirty{}: open-source plug-and-play demodulator/decoder for 1030MHz uplink}

\date{}


\author{
\IEEEauthorblockN{Lassi Laaksosaari$^\circledast$~\thanks{$^\circledast$ Equally contributing authors.}, 
Hannu Turtianen$^\circledast$, 
Syed Khandker, 
Andrei Costin $^\ddagger$~\thanks{$^\ddagger$ Corresponding author: ancostin@jyu.fi}}
\IEEEauthorblockA{
Faculty of Information Technology $^\ast$ \\
University of Jyv\"askyl\"a \\ 
Jyv\"askyl\"a, Finland \\ 
\{laaklavy, turthzu, syibkhan, ancostin\}@jyu.fi\\
\thanks{$^\ast$ All authors are affiliated with: Faculty of Information Technology, University of Jyv\"askyl\"a, Finland. }
\thanks{AUTHORS’ PREPRINT, DRAFT ACCEPTED BY REVIEWERS/EDITORS of IEEE Aerospace and Electronic Systems Magazine (AESM) on 13th Feb 2023.}
}
}

\maketitle

\begin{abstract}
Automatic Dependent Surveillance (ADS), Automatic Dependent Surveillance-Broadcast (ADS-B), Secondary Surveillance Radars (SSR), and Mode S are key air surveillance technologies representing a critical component of next-generation air transportation systems. 
%
However, compared to 1090MHz demodulators and decoders, which have plenty of implementations, the 1030MHz uplink receivers are, in general, scarcely, if at all, represented. 

In this paper, we present the development and evaluation of \dumptenthirty{} -- cross-platform plug-and-play open-source implementation for decoding 1030MHz uplink Mode A/C/S interrogations. 
We demonstrate and detail an agile development process of building \dumptenthirty{} by adapting a state-of-the-art dump1090 design and implementation. 
In our repeated experiments, \dumptenthirty{} achieves a high detection accuracy of 1030MHz interrogation signals based on lab evaluation using synthetically-generated interrogation signals. 
We also discuss a handful of practical use cases where \dumptenthirty{} can find immediate application and implementation, both in research and industrial settings. 
%
\end{abstract}



\section{Introduction}
\label{sec:intro}

Automatic Dependent Surveillance (ADS), Automatic Dependent Surveillance-Broadcast (ADS-B), and Mode S are key air surveillance technologies representing a critical component of next-generation air transportation systems. 
These technologies significantly simplify aircraft surveillance technology and aim to improve airborne traffic situational awareness. 
Based on the system's roles, Mode S and ADS/ADS-B have two main components.
One component is the \emph{interrogators}, such as Secondary Surveillance Radars (SSR), Air Traffic Controllers (ATC), and aircraft/vehicles, which operate at 1030MHz. 
The other component is the \emph{responders}, such as ADS-B/Mode S equipped aircraft, vehicles, and ground infrastructure, which operate at 1090MHz. 
The most exciting information about an ADS-B vehicle (e.g., ICAO24 aircraft address, flight number, GPS location) is in 1090MHz data from the responding transponders. 
This 1090MHz data can be automatically and periodically broadcasted (e.g., ADS-B) or replied to by a responder based on specific interrogation requests over 1030MHz from an interrogator. 

In recent years, the main focus of the media, as well as the open-source and commercial projects, was on ADS-B data over 1090MHz, 
which was mainly driven by online services such as \url{www.flightradar24.com}, cybersecurity research revelations 
such as ``Ghost in the Air(Traffic)''~\cite{costin2012ghost,khandker2021cybersecurity,khandker2022cybersecurity-adsb,juvonen2022log4j}, and projects such as dump1090~\cite{sanfilippo2013dump1090}. 
Nevertheless, the interrogation data over 1030MHz also presents valuable information (especially for Machine Learning driven aviation deployments) that researchers and practitioners miss. 

Compared to 1090MHz ADS-B demodulators and decoders, which have plenty of implementations~\cite{sanfilippo2013dump1090} and forks~\cite{jowettgithub}, the 1030MHz ADS-B decoders are, in general, scarcely, if at all, represented. 
We are aware of only a handful of ADS-B uplink 1030MHz demodulators/decoders.
The RTL1030~\cite{rtl1030} is a \emph{proprietary closed-source} implementation of a 1030MHz decoder that \emph{runs only on Windows}. 
The OpenSky project~\cite{strohmeier2015opensky} provides a Matlab-based implementation of uplink 1030MHz demodulator and decoder~\footnote{\url{https://github.com/openskynetwork/modes-uplink-decoder}}. However, it has several limitations. It only implements \emph{offline mode} demodulation and decoding based on pre-captured samples. Also, it mainly addresses the \emph{decoding} of pure Mode S 1030MHz uplink interrogations and does not handle other 1030MHz uplink signals (e.g., Mode A/C, Mode A/C All Call). Finally, even if adapted to decode in real-time, installing and running Matlab on embedded devices (e.g., Raspberry Pi, specialized ADS-B decoders) is unnecessary and impractical overkill. 
The 1090 Megahertz Riddle's pyModeS~\cite{sun20211090} provides a Python-based implementation of uplink demodulator and decoder~\footnote{\url{https://github.com/junzis/pyModeS/blob/master/pyModeS/decoder/uplink.py}}. However, it has several limitations. Similar to OpenSky's one, it only implements \emph{offline mode} and mainly addresses the \emph{decoding} of pure Mode S uplink interrogation. 

Therefore, a core motivation behind this work is to bring to the research and practitioner communities a cross-platform, open-source, and plug-and-plug (similar to dump1090) alternative to any limited or proprietary implementations used in air traffic control (ATC), air traffic management (ATM), and crowd-sourcing deployments. 
More specifically, we aimed that \dumptenthirty{} offers \emph{real-time} demodulation and decoding, and to be \emph{cross-platform plug-and-play} similar to dump1090. These features would allow to click-and-run \dumptenthirty{} along dump1090 on most common devices (such as Raspberry Pi).

Our contributions to this work are as follows:
\begin{enumerate}
\item We develop, evaluate, and present \dumptenthirty{} -- cross-platform plug-and-play open-source implementation for processing 1030MHz uplink Mode A/C/S interrogations.

\item We propose and discuss a handful of practical use cases where \dumptenthirty{} can find immediate application and implementation, both in research and industrial settings.

\item For evaluation and further improvements, we release (upon peer-review acceptance) the relevant artifacts (e.g., code, data, documentation, examples) as open-source: 
\url{https://github.com/Fuziih/dump1030}. 

\item We further invite the community to support the open-source evolution by adding to Mode S detection also \emph{real-time decoding capabilities} for Mode S P6 datablock (see Section~\ref{sec:sample-limit}). 
\end{enumerate}

The rest of this paper is organized as follows.
In Section~\ref{sec:background} we present some background on 1030MHz aviation-related technologies. 
In Section~\ref{sec:impl} we detail our implementation along with the evaluation results. 
We further motivate the applicability of \dumptenthirty{} by discussing immediate use-cases in Section~\ref{sec:use-cases}. 
We present the related work in Section~\ref{sec:relwork}. 
Finally, we conclude the paper with Section~\ref{sec:concl}.


\section{Background}
\label{sec:background}

\begin{figure}[htb]
\centering
\includegraphics[width = 1.0\columnwidth]{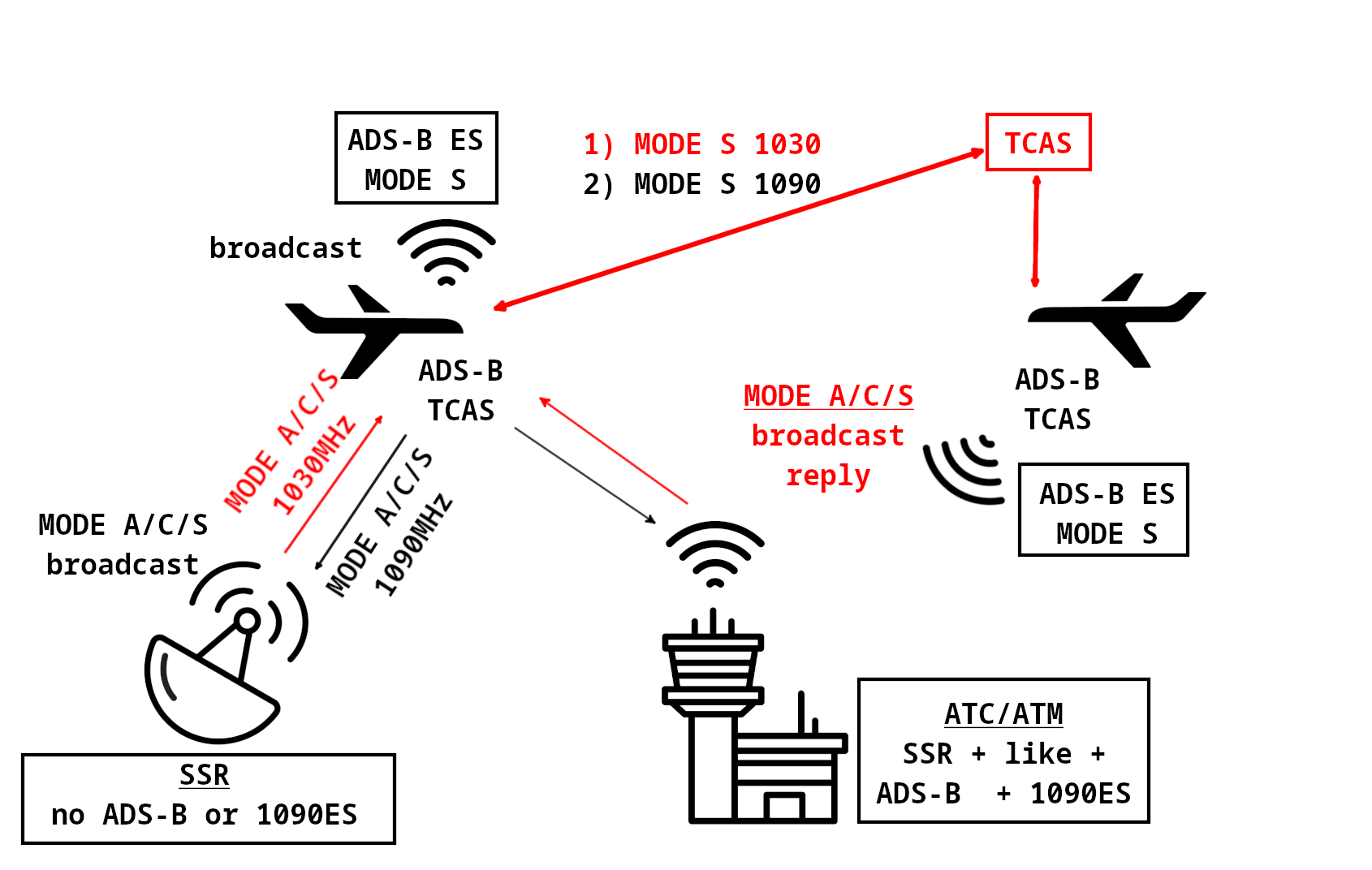}
\caption{Simplified diagram of 1030/1090 aviation communication environment.}
\label{fig:1030_signals}
\end{figure}

In this section, we briefly introduce aircraft surveillance technology and protocols. 
Fig.~\ref{fig:1030_signals} depicts a simplified view of the 1030/1090 aviation communication environment.

\subsection{Mode A/C/S Background}
\label{sec:background-mode}

\begin{figure*}[htb]
\centering
\includegraphics[width = 0.95\columnwidth, trim = {1.5in 4.50in 1.3in 1.0in}]{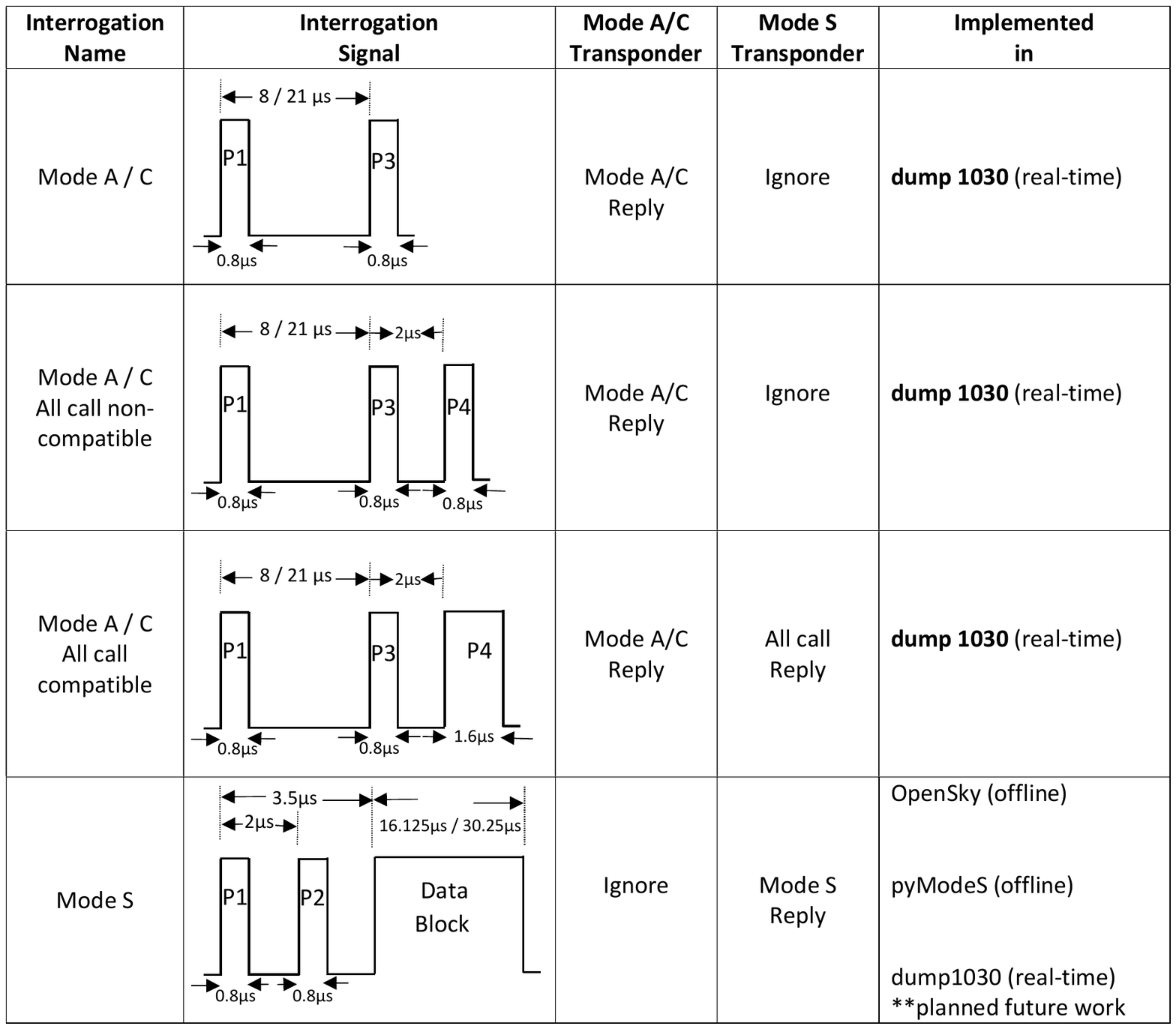}
\caption{Types of radio frequency signals employed for various 1030MHz uplink interrogations.}
\label{fig:1030types}
\end{figure*}

Aircraft surveillance relied only on the primary surveillance radar (PSR) in the early days. With PSR, the aircraft's position is measured by distance and angle to the radar, but it only indicates the aircraft's position and does not identify them. However, identifying an aircraft is very important, e.g., when distinguishing a friend from a foe or when guiding optimal routing is required. The secondary surveillance radar (SSR) came into use to identify an aircraft correctly. The SSR transmits interrogation pulses at 1030MHz, known as Mode A and C. The Mode A and Mode C pulsing allow the SSR to interrogate the aircraft's identity and barometric altitude continuously. The reply to the interrogation is given at 1090MHz. When many aircraft share the same air space, it is ambiguous for which aircraft an interrogation is made. The idea of Mode S (or selective interrogation) evolved to make the scenario clear. Each Mode S transponder-equipped aircraft is assigned a unique address code called ICAO24 code, by which interrogations can be directed to a particular aircraft, and replies can be unambiguously identified. The transponders recognize interrogation pulses based on the shape of the pulse. Fig.~\ref{fig:1030types} shows the types of interrogation and supportive transponder. 

\subsection{ADS-B Background}
\label{sec:background-adsb}

ADS-B is an air surveillance technology that periodically transmits an aircraft's position, identity, velocity, and traffic-relation information to other aircraft via the radio link. The most important part of the ADS-B data is the precise location information generally determined by global navigation satellite systems (GNSS) such as GPS. Unlike Mode A or C, ADS-B technology does not require any interrogation from ATC, as it works automatically. Based on the physical layer used for Radio Frequency (RF) communications, there are two ADS-B types -- 1090MHz Extended Squitter (1090ES) and 978MHz universal access transceiver (UAT978). ADS-B 1090ES works worldwide on 1090MHz, while ADS-B UAT978 works mainly in the USA and is mandated for general aviation flying under 18,000 feet.  

\subsection{TCAS Background}
\label{sec:background-tcas}

A traffic collision avoidance system (TCAS) is an aircraft collision avoidance system designed to reduce the incidence of a mid-air collision between aircraft. A Mode S transponder is required for TCAS. The TCAS unit sends Mode S interrogations (Uplink Format 0, or UF0) to get other aircraft's range, bearing, and altitude, while further aircraft replies with (Downlink Format 0) containing its altitude. The closest point of approach (CPA) and time until the CPA (TAU) are calculated from the acquired data. According to International Civil Aviation Organization (ICAO) instrument flight rules, aircraft flying below 29,000 feet should maintain a vertical separation of no less than 1,000 feet of altitude. Above 29,000 feet, the separation needs to be 2,000 feet or greater. If TAU is about 40 seconds, a possible collision alert threat, called Traffic Advisory (TA), will be triggered. If TAU is about 25 seconds, a real threat of collision or Resolution Advisory (RA) will be triggered. In this instance, TCAS will automatically initiate a mutual avoidance maneuver for both aircraft.


\section{Implementation and Evaluation}
\label{sec:impl}

\begin{figure}[htb]
\centering
\includegraphics[width = 1.05\columnwidth, trim = {0.0in 0.0in 0.0in 0.0in,,clip=true}]{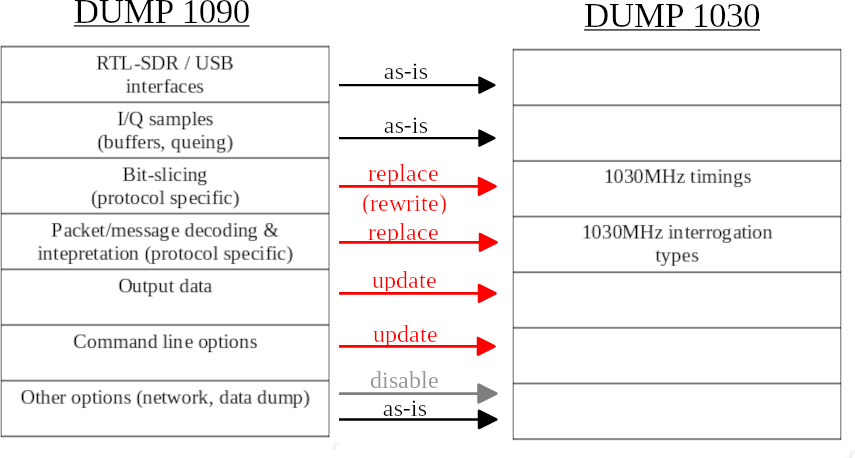}
\caption{Design process of agile implementation of \dumptenthirty{} starting from dump1090.}
\label{fig:1090to1030}
\end{figure}

We started implementing \dumptenthirty{} by taking  antirez's implementation of dump1090~\cite{sanfilippo2013dump1090} as a base. The dump1090 software is perhaps one of the most popular Mode S decoders for 1090MHz downlink messages. The \dumptenthirty{} design and code structure are otherwise very similar, but it does not require decoding the ADS-B 1090ES data. It also does not require additional input features such as receiving data from a network client, which we did not include in \dumptenthirty{}. 
In Fig.~\ref{fig:1090to1030}, we present the high-level approach to agile design and development practices that allowed us to adapt dump1090 into \dumptenthirty{} quickly. 
Our overall experience suggests that similar techniques could and should be used for quick prototyping other similar RF-related protocols in aviation, maritime, aerospace, and satellite communications. 

\begin{figure}[htb]
\centering
\includegraphics[width = 0.8\columnwidth]{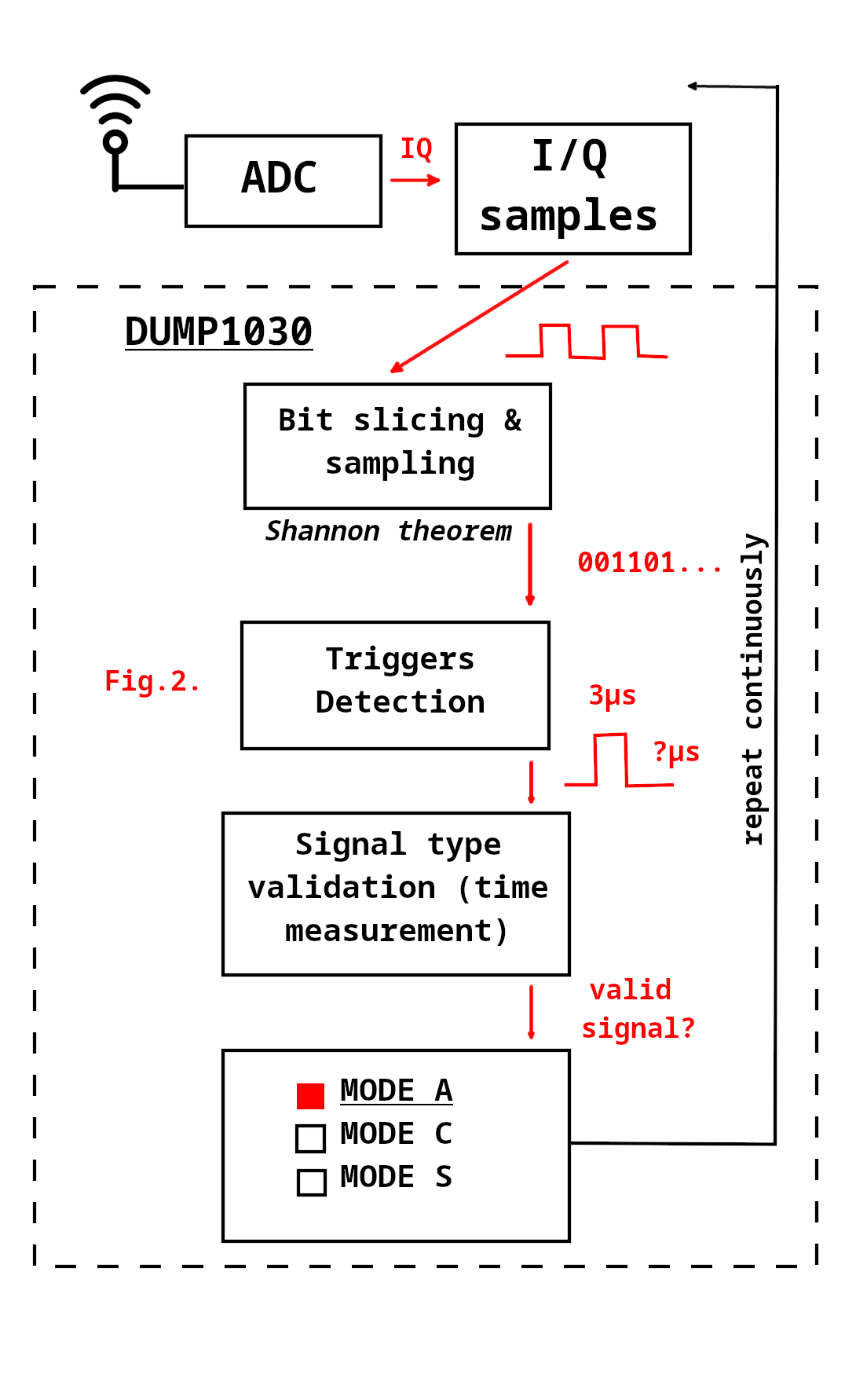}
\caption{Logical operation flowchart of \dumptenthirty{}. }
\label{fig:dump1030}
\end{figure}

In Fig.~\ref{fig:dump1030}, we present the high-level logical flowchart of main loop actions performed by \dumptenthirty{} during its operation, and below we further detail some of the aspects of the operation. 

The \dumptenthirty{} translates 8-bit In-Phase and Quadrature (IQ) values ranging from -128 to +128 to non-negative values ranging from 0 to 255. Checking a message starts by comparing P1 with values that are not pulse values in any SSR interrogation before a P3 pulse, as the P1 pulse exists at the same place and size in every SSR interrogation. If the P1 test passes, we check the rest of the message for the Mode S uplink preamble because Mode S preamble is the shortest and therefore requires the least number of comparisons which in turn requires the least computational processing. If the Mode S check passes, we add one message to statistics and skip 49 samples. Forty-nine samples are 19.6~$\mu$s at a sample rate of 2.5 MSPS. 19.625~$\mu$s is the full length of the short (56 bits) Mode S message, hence why we skip 49 samples. If Mode S check fails, we continue to Mode A check since it is shorter than the Mode C check. If the Mode A check fails, we continue to the Mode C check, and after the Mode C check fails, we start the checking loop again from the following sample. If passing the Mode A or Mode C check, the program skips to the P4 check. 
The P4 check starts by checking for short P4 and, upon failing, continues to long P4 check. If both P4 checks fail, the program marks the message as standard Mode A/C and skips the corresponding number of samples based on the message type. If either of the P4 checks pass, the aforementioned actions are done based on the message type. Every check fails if at least one of the comparisons does not pass. The comparisons are made by comparing absolute difference and relative difference of amplitude. The absolute difference is measured by adding a user-configurable parameter to the expected non-pulse sample and comparing it to the expected pulse sample. Relative difference is measured by dividing the expected non-pulse sample by the expected pulse sample and comparing it to the user-configurable parameter.
In most cases, relative difference comparison detects correct patterns matching message types, while absolute difference ensures that random noise is not recognized as a message in low noise areas. P4 check does not use absolute difference comparisons since earlier checks already incorporate checking absolute; therefore, the expectation is that it is not required. P4 check mainly checks that non-pulse values not next to pulse samples are less than half amplitude to P4 pulse samples on default values. \dumptenthirty{} also includes the option to set fixed amplitude filters for non-pulse values not next to pulse samples, non-pulse values next to pulse samples, and pulse samples. 
%
\dumptenthirty{} can output recommended values for these fixed variables based on averages of successfully accepted messages during a particular running period. 
These features can be set to determine borderline filter values on top of the ones based on comparing values. 
Pulse means consecutive values interpreted as one (1) by code in a message. 
Pulse break means consecutive values interpreted as zero (0) by code in a message. 
We use these terms to make the text more understandable and concise.

To detect every pulse properly, the time between samples must divide evenly with every pulse width and time between every pulse. If the time between samples exceeds the smallest pulse width or pulse break width, the receiver might miss the most negligible pulse or pulse breaks entirely. If the time between samples does not divide evenly with pulse break width or pulse width, but the sample rate is lower than the smallest pulse width or pulse break, the pulse or pulse break can be read in length one sample longer or shorter. This depends on which part of the message receiver takes the first sample. Mode S 1090MHz downlink messages preamble have a pulse width of 0.5 microseconds. We are not interested in pulse width or pulse break width of Mode S uplink data block because our hardware cannot read them properly, as demonstrated in later section~\ref{sec:sample-limit}. 
The 1030MHz uplink messages have a pulse width of 0.8~$\mu$s and 1.6~$\mu$s. Widths of pulse breaks in uplink messages are 0.7, 1.2, 7.2, and 20.2~$\mu$s. The greatest common divisor for the numbers above is 0.1~$\mu$s or 10 MSPS which is way more than most RTL-SDR radios are capable of. Taking a sample every 0.4~$\mu$s or, in other words, 2.5 MSPS seems to be the best solution because only pulse breaks or pulses that are not divided evenly are 20.2 and 0.7~$\mu$s. Differences in being evenly divided are 0.2~$\mu$s and 0.1~$\mu$s, respectively, which are just within the error marginal standards set by ICAO~\cite{srvsop-icao}. The cheapest RTL-SDR radios were already creating errors at 2.5 MSPS, and these hardware limitations will be discussed more in detail in section~\ref{sec:sample-limit}. Fig.~\ref{fig:dump1030} details the \dumptenthirty{} normal operation.

\subsection{Evaluation Results}
\label{sec:eval}

\subsubsection{Lab Evaluation}
\label{sec:eval-lab}

During the development and testing of \dumptenthirty{}, we have used the following best practices:
\begin{itemize}
\item We configure the transmitters (e.g., HackRF) and receivers (e.g., RTL-SDR) to use the ISM-band, meaning that the transmission and reception of the signal waves were done on the central carrier frequency of 433.800\,MHz. 
In Finland, the 432--438\,MHz ISM-band is allocated for transceivers that are exempt from licensing~\cite{traficom2021}. 

\item We also set the lowest transmit power to limit unintended interference in the unlicensed ISM band. 

\item In addition, we use a certified ``faraday cage'' --- specifically a Disklabs Faraday Bag --- featuring a double layer military-grade RF faraday shielding, which is also commonly used for well-contained wireless and RF testing and forensics. 

\item Moreover, we use a certified radio power density meter --- specifically a TriField Model TF2 EMF Meter --- to double-check and ensure that the signals do not escape the faraday cage/lab premises. 
\end{itemize}
All these precaution measures are complementary and ensure a well-controlled environment, which is also in line with commonly accepted practices~\cite{khandker2021cybersecurity,khandker2022cybersecurity-adsb,strohmeier2022building,khandker2022cybersecurity-ais,juvonen2022log4j}.

\begin{figure}[htb]
\centering
\includegraphics[width = 1.00\columnwidth, trim = {0.0in 0.0in 0.0in 0.0in}]{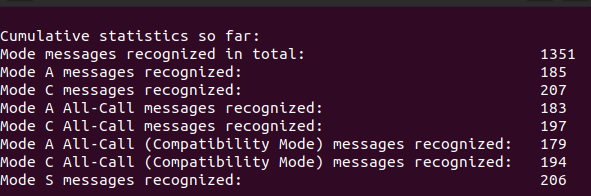}
\caption{Example of running \dumptenthirty{} in statistics mode.}
\label{fig:dump1030-in-action}
\end{figure}

During the development, we tested \dumptenthirty{} by sending seven types of perfect (i.e., noiseless) SSR interrogations by a transmission-enabled software-defined radio called HackRF~\cite{hackrf}.
At first, through the Python program, we created the byte order of the interrogation pulse. 
The byte order or payload was fed into a program called ``GNU Radio Companion'' (GRC)~\cite{grc}~\footnote{GNU stands for ``GNU's Not Unix'', and is an extensive collection of free software -- \url{https://www.gnu.org}.} to produce the IQ of the signal. 
Later, using HackRF, the IQ were transmitted to the \dumptenthirty{}-equipped receiver. 
In the Fig.~\ref{fig:dump1030-in-action} we present a sample screenshot of the \dumptenthirty{} in action. 

The sample rate defines the pulse duration. The 2.5 MSPS resulted in a duration of 0.4~$\mu$s for each sample. 
The difference between the start of the message and the start of the P3 pulse defines whether the message is Mode C or Mode A. In Mode C message time difference between P1 and P3 is 21~$\mu$s while in Mode A message, it is 8~$\mu$s. 
The Mode A and Mode C messages have the three interrogation sub-categories as depicted in Fig.~\ref{fig:1030types} (the first three content rows from top to bottom). 
From the top of Fig.~\ref{fig:1030types}; the first interrogation type depicts a Mode A/C standard message, the second shows a Mode A/C all-call with a short P4 pulse, the third depicts a Mode A/C all-call (Compatibility Mode) with a long P4 pulse, and the fourth type represents a Mode S message including the P6 data block.
\dumptenthirty{} error rates are partially dependent on user configurable parameters as explained in later section~\ref{sec:chal-limit}. 
As of this writing, we are running experiments to test how many messages \dumptenthirty{} failed to detect completely under different configurations in order to suggest future users best parameters for various use-cases, and we leave reporting these experiments as a future work. 
Only wrong messages were recorded since that was the goal of the research. 
It is expected that missing messages depend highly on how strict or, in other words, low configurable parameters are; therefore, the user can control the balance between false positives and missed messages. As seen in Table~\ref{tab:falsely_detection}, highest false detection percentages are in Mode A/C all-call non-compatible.

\begin{table}[!h]
 \caption{Confusion matrix for \dumptenthirty{} 1030MHz signal detection during lab ``ground truth'' evaluation.}
 \label{tab:falsely_detection}
\resizebox{\columnwidth}{!}{
\begin{tabular}{@{}|l|l|@{}}
\toprule
\multicolumn{1}{|c|}{\textbf{Interrogation type sent}}                     & \multicolumn{1}{c|}{\textbf{Interrogation type received (false detection ratio)}} \\ \midrule
Mode A                                          & No false detection                               \\ \midrule
\multirow{2}{*}{Mode A all call compatible}   & Mode A 0.17\%                                   \\ \cmidrule(l){2-2} 
                                                & Mode A all call non-compatible 0.02\%           \\ \midrule
Mode A all call non-compatible                  & Mode S 1.87\%                                   \\ \midrule
Mode C                                          & Mode C all call non-compatible 0.01\%           \\ \midrule
\multirow{2}{*}{Mode C all call compatible}    & Mode C 0.05\%                                   \\ \cmidrule(l){2-2} 
                                                & Mode C all call non-compatible 0.03\%           \\ \midrule
\multirow{2}{*}{Mode C all call non-compatible} & Mode C 0.09\%                                   \\ \cmidrule(l){2-2} 
                                                & Mode S 7.05\%                                   \\ \midrule
Mode S                                          & No false detection                               \\ \bottomrule
\end{tabular}
}
\end{table}

This behavior is a consequence of Mode A/C all-call non-compatible messages having identical sections starting from the start of P3 compared to Mode S preamble (shown in Fig.~\ref{fig:1030types}).
Mode A/C all-call (Compatibility mode) and Mode A/C all-call messages can be misinterpreted as standard Mode A/C messages because in the case of both P4 checks failing \dumptenthirty{} marks message always as standard Mode A/C message. The rest of the errors are relatively low (0.01~\%, 0.02~\%, and 0.03~\%, respectively). These errors can be assumed to occur due to hardware limitations and other factors like noise. Section Challenges and Limitations~\ref{sec:chal-limit} explains in more detail why these errors happen and what challenges \dumptenthirty{} has when detecting messages.

\subsection{Challenges and Limitations}
\label{sec:chal-limit}

This section presents some of the challenges we faced and some of the limitations of developing or using \dumptenthirty{}. 

\subsubsection{Real-world RF and noise}

\begin{figure}[!b]
\centering
\includegraphics[width = 1.0\columnwidth, trim = {0.0in 0.0in 0.0in 0.0in}]{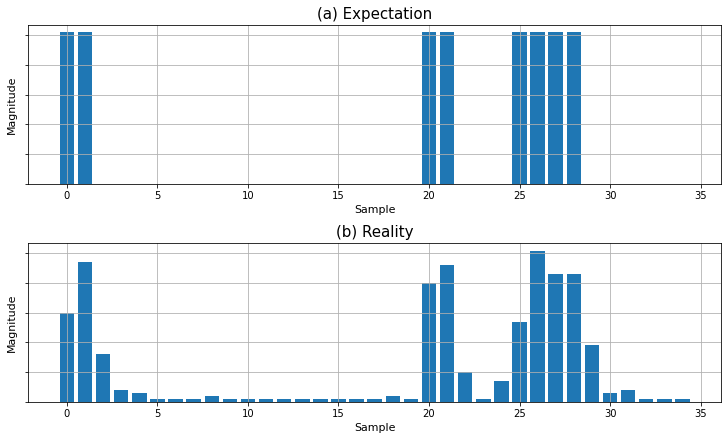}
\caption{Challenges with signal and pulse shapes -- example of ``Mode A All-Call'' with ideal expected signal vs. actual signal received in the lab transmit-receive experiments.}
\label{fig:exp}
\end{figure}

\dumptenthirty{}'s aim is to detect as weak as possible SSR-interrogation pulse patterns while keeping the number of false positives as low as possible. Fig~\ref{fig:exp}(a) shows a theoretical Mode A all-call (compatibility mode) signal.
Fig.~\ref{fig:exp}(b) presents the same signal at the receiver end after being transmitted by HackRF and received by RTL-SDR. 
The X-axis represents samples, while the y-axis shows the magnitude of the samples. Due to the sample rate of 2.5 MSPS, each sample's duration is 0.4~$\mu$s. Therefore, two samples are needed to express the P1 pulse, which is 0.8~$\mu$s. In total, 29 samples were required to represent the whole Mode A all-call (compatibility mode) signal.  
Now, the transmitted signal gets affected by many natural phenomena such as reflection, refraction, diffraction, absorption, and attenuation. Moreover, in analog to digital or vice versa, conversion previous sample's energy could affect the following sample, or if the SDR is not very quick in sampling/switching, it may result in different magnitudes in the received samples. Therefore, in Fig~\ref{fig:exp}(b), we can see some unwanted noise in the receiver end. Generally, by setting some threshold and logic, the noise is eliminated.  

During testing, we examined that expected non-pulse samples close to expected P4 pulse samples can be as high or even higher than expected P4 pulse samples. In addition, expected non-P4 pulse samples not next to expected P4 pulse samples were also examined to be higher than their counterparts measured before the P3 pulse. Therefore expected non-pulse samples are divided into four groups based on their behavior: expected non-pulse samples that are not next to expected pulse samples between P1 and P3, expected non-pulse samples that are right next to expected pulse samples between P1 and P3, and two aforementioned separate groups after P3 (also referred in the code as \emph{``P4 check variables''}, e.g., \emph{diffratiop4}, \emph{diffratioclosep4}). All these four groups have their own separate user configurable parameters. Mode A/C all-call and Mode A/C all-call (Compatibility mode) are only 0.8~$\mu$s longer P4 pulse, which equals two samples of difference at the sample rate of 2.5 MSPS. Since the first sample after the end of the P4 pulse can be even higher than the P4 pulse sample, only one sample determines if a message is Mode A/C all-call or Mode A/C all-call (Compatibility mode) in such scenarios.
Consequently, based on user-configurable parameters (those that handle the P4 detection and threshold checks), more strict parameters lead to the fact that certain messages, especially Mode A/C all-call and Mode A/C all-call (Compatibility mode), to be misinterpreted as standard Mode A/C messages when both P4 threshold checks fail. However, by setting \emph{``P4 check variables''} too high, \dumptenthirty{} may start to misinterpret Mode A/C all-call (Compatibility mode) messages as Mode A/C all-call messages since the difference between these two message types is only two samples, and Mode A/C all-call P4 check is first of the two. These misinterpretation problems could be mitigated by further separating user configurable parameters so that two of the samples above can have separate configurable comparison variables and other P4 samples with different configurable comparison variables. However, our testing achieved relatively low error rates ranging from 0.02--0.17~\%. Therefore we leave such further variables fine-tuning as future work. 
In addition, another interpretation problem is present when Mode S preamble is identical to part of the Mode A/C all-call message. In turn, if the Mode A/C check fails on Mode A/C all-call message, the program will loop until it reaches the start of P3 and marks it as a Mode S message. However, Mode S check uses stricter variables by default than the P4 check, which can mitigate the amount of false detection in this scenario. Furthermore, Mode C all-call has significantly more false detection (7.05~\%) than Mode A all-call (1.87~\%) because the failure of detecting Mode C message is much more likely due to longer length, which in turn leads to more comparisons. Comparing values from the start of the pulse P6 data block could fix this misinterpretation since there is a continuous 1.25~$\mu$s pulse~\cite{Bredemeyer,Mateu}, and we leave this marginal improvement for future work.

\subsubsection{Sampling-rate limitations of RTL-SDR}
\label{sec:sample-limit}

RTL-SDR dongles have a theoretical maximum sample rate of 3.2 MS/s; however, in practice, most dongles start dropping samples after passing 2.4 MS/s~\cite{9092398,8356821}. Testing was conducted successfully with RTL2832U, while the FC0012 equipped dongle repeatedly failed with \dumptenthirty{} at the required sample rate of 2.5 MS/s. 
For more detailed introduction to RTL-SDR dongles, their types as well as their practical applications and constraints, we reference the interested readers to relevant works~\cite{danymol2013real,sruthi2013low,aguilar2020comparative}. 

The Mode S interrogation requires a minimum sample rate of 2 MS/s due to preamble pulse width of 0.5~$\mu$s and P6 data block of 56 bits or 112 bits being encoded in pulse position modulation throughout 56~$\mu$s or 112~$\mu$s respectively. Pulse position modulation uses one pulse and one non-pulse to represent every bit; therefore, a 112-bit message over 112 $\mu$s leads to a pulse/non-pulse width of 0.5~$\mu$s. The P6 data block of Mode S uplink message differs in two ways from the P6 data block of Mode S downlink message: it uses Differential phase shift keying (DPSK)  and its length in time is 30.25~$\mu$s or 16.125~$\mu$s. DPSK requires one pulse per bit; thus, 112 bits over 30.25~$\mu$s leads to a pulse width of 0.25~$\mu$s~\cite{Bredemeyer,Mateu}. The pulse width of 0.25~$\mu$s would require a minimum sample rate of 4 MS/s which is beyond the theoretical maximum sample rate of RTL-SDR devices. 
Hence, decoding the P6 data block of Mode S uplink messages (similar to what offline decoders from OpenSky and pyModeS offer -- see notes in Section~\ref{sec:intro}) requires more capable SDR receivers (e.g., HackRF or BladeRF) and a non-trivial improvement of the \dumptenthirty{} software; therefore, we leave this as future work. 

\section{Use-Case Scenarios}
\label{sec:use-cases}

\subsection{Improved Big-Data Analytics of ADS-B Communications}

Projects such as OpenSky~\cite{strohmeier2015opensky}, FlightRadar24~\cite{flightradar} collect ADS-B data globally, 
and then store and process that data for multiple purposes such as research reports and analytics~\cite{schafer2016opensky,schafer2019opensky}. 
To the best of our knowledge, unfortunately, at present, all such (crowd-source) systems collect only downlink messages (e.g., 1090MHz responses)~\cite{schafer2019opensky}. 
This means the collected data may lack complete or accurate context, mainly when it aims for TCAS and SSR interrogations analysis~\cite{schafer2016opensky,schafer2019opensky}. 
Therefore, tools such as \dumptenthirty{} can bring essential improvement and granularity to the ADS-B data collection of paired and contextualized 1030MHz (uplink) and 1090MHz (downlink) transmissions. 
Moreover, \dumptenthirty{} can also help build more advanced and mature ML models considering more communication, such as 1030MHz uplink messages.

\subsection{Towards Open-Source TCAS Implementations}

\begin{figure}[htb]
\centering
\includegraphics[width = 1.10\columnwidth]{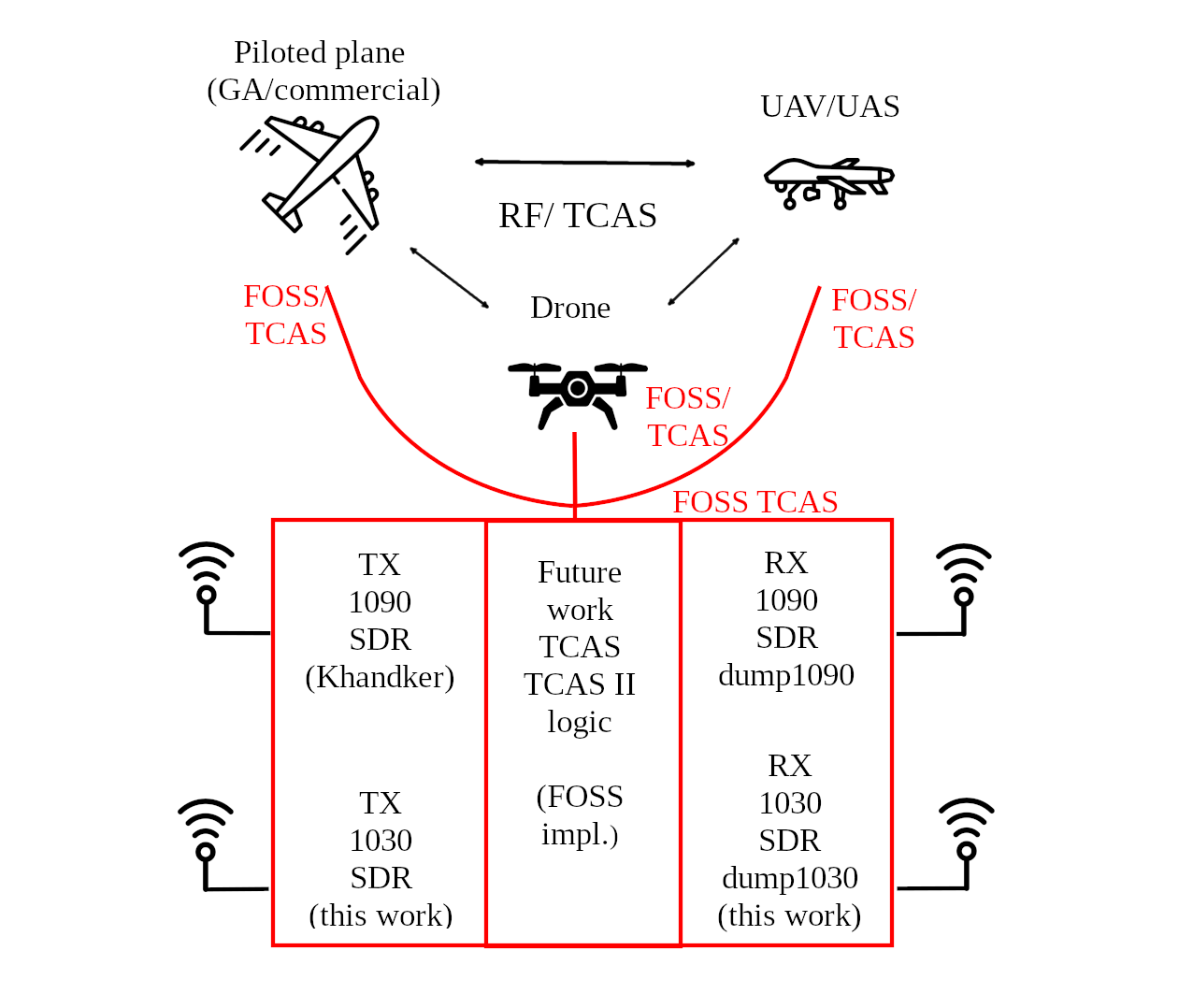}
\caption{Conceptual design of different vehicles having onboard TCAS implementation based on open-source projects such as dump1090 and \dumptenthirty{}. The design assumes the depicted FOSS/TCAS box is present in all airborne vehicles.}
\label{fig:1030-TCAS}
\end{figure}

The 1030MHz interrogations are an essential part of the Traffic Collision-Avoidance System (TCAS)~\cite{harman1996beacon,panken2012measurements}.
Therefore, to move forward toward open-source TCAS implementations, whether passive monitoring-only receivers or active transceiver systems, it is essential to have open-source implementations able to receive the 1030MHz interrogations at least.
In Fig.~\ref{fig:1030-TCAS}, we present a high-level conceptual model of achieving, implementing, and using free and open-source (FOSS) implementation of TCAS. 

\subsection{Improved Localization of 1030MHz Stations and SSRs}

\begin{figure}[htb]
\centering
\includegraphics[width = 1.10\columnwidth]{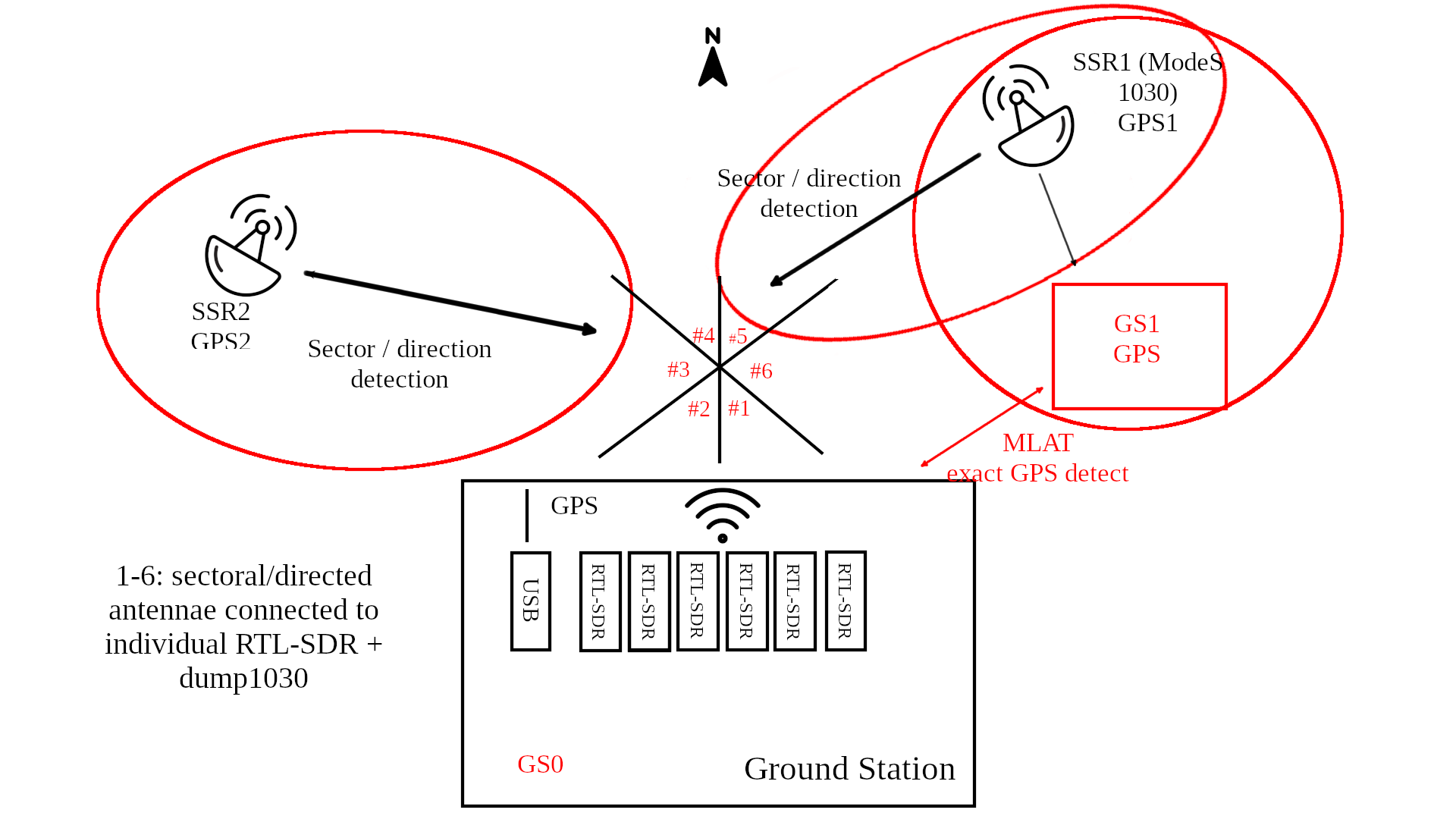}
\caption{Conceptual design using single-station (GS0) or multiple-station (GS0+GS1) deployments of \dumptenthirty{} for finding direction (SSR1) and exact GPS location (SSR2) of 1030MHz interrogators using sectoral antennas and Multilateration MLAT (GS0+GS1) -- in this example in the center is depicted a 6-sector antenna of GS0, and depiction of GS1 is minimized.}
\label{fig:1030-MLAT}
\end{figure}

\dumptenthirty{} coupled with multiple RTL-SDR connected to sector-directed array antenna (e.g., six sectors 60 degrees each) could be an effective tool to detect the exact location of ``hidden'' SSR stations. 
Such ``hidden'' SSR stations could be either clandestine or belong to a military adversary; therefore, their detection could be helpful in certain situations. 
Moreover, several ground stations (GS), implemented on Raspberry Pi or laptops, could be used essentially to perform multilateration (MLAT) on 1030MHz signals, hence accurately pinpointing the location of such 1030MHz SSR interrogation stations. 
In Fig.~\ref{fig:1030-MLAT} we present a high-level conceptual model on how single and multiple ground stations (GS) (coupled with GPS and numerous sectoral antennas connected to individual \dumptenthirty{} instances) can be used to achieve precise localization of 1030MHz SSR interrogation stations. 

To demonstrate and evaluate this approach in practice, we have tried to acquire a 6-sector antenna supporting Mode S bands 
1030MHz and 1090Mhz~\footnote{The ``6 Sector Squitter, Ground Station Antenna Part Number: 530300-100'' -- \url{https://www.dbsant.com/antennas/6-sector-squitter-ground-station-antenna/} -- costs around 60.000 USD.}, 
however it proved to be prohibitively expensive for a research lab experiment. 
Therefore, the current pricing and limited offerings prompt further research on open-hardware and open-firmware active antennas that could potentially allow research labs such as ours to build these antennae at a price range of several hundred and no more than a couple of thousand USD/EUR.

\subsection{Compliance Checks of 1090MHz Transponders}

As part of \dumptenthirty{} project, besides the source codes, we also release \emph{reference 1030MHz interrogation signals}.
In very well-controlled labs (such as detailed in~\cite{strohmeier2022building}), the researchers can send the interrogation signals to some particular 1090MHz transponder under test, 
and then observe its compliance to functional (i.e., responds or not), performance (e.g., response within required timeslot, is response signal correct and compliant), and cybersecurity~\cite{cohen2019security}, requirements. 
Moreover, \dumptenthirty{} can play the receiving end of the response to a 1030MHz interrogation, thus assisting the compliance test and validation, both from functional, operational, and cybersecurity perspectives. 
For example, the inclusion of \dumptenthirty{} into an avionics test lab such as the one proposed by Strohmeier et al.~\cite{strohmeier2022building} could be immediate and straightforward. 
In summary, in this context our \dumptenthirty{} project (including \emph{reference 1030MHz interrogation signals}) is a valuable all-in-one 1030MHz toolset. 

\subsection{Applications in Low-Cost Drones and UAV/UAS Systems}

Industry-grade aerospace vehicles and aircraft will likely rely on commercial-grade and certified implementations of 1030MHz receivers and decoders (e.g., certified TCAS). 
However, such TCAS systems cost tens (and hundreds) of thousands of EUR/USD. 
Therefore, we believe \dumptenthirty{} can be instrumental in the prototyping, development, and commercialization of low-cost drones, UAVs, and UAS systems, especially those without stringent functional or performance requirements. 
Being released as open-source and under a permissive license, it will allow faster prototyping and commercialization of low-cost drones, UAVs, UAS systems, 
where the time-to-market and cost of certification for existing/standard 1030MHz implementations can be prohibitive, so \dumptenthirty{} can prove an essential (and only) alternative.


\section{Related work}
\label{sec:relwork}

Cohen and Smith implemented a low-cost SSR/ADS-B decoder~\cite{cohen2implementation}. 
They showed that aircraft ascent or descent, response to ground interrogation, TCAS Resolution Advisory, etc., could be decoded with their portable setup. 
They practically tested the setup near the JFK airport in New York, and their main goal was to develop a cost-effective system for SSR. 
Salvatore Sanfilippo initially released dump1090, a Mode S decoder for RTL-SDR devices, back in 2013~\cite{sanfilippo2013dump1090}. dump1090 decodes all downlink Mode S message types except df24, which is rarely used in the real world. In addition, dump1090 included network streaming capabilities and the ability to show aircraft interactively on a list or Google maps. Multiple forks have evolved from dump1090. dump1090-fa, updated by Flightaware, is probably the most regularly updated and feature-rich fork. While all the basic functionalities have remained the same, dump1090-fa~\cite{jowettgithub} includes installation through Debian/Raspbian packages and support for LimeSDR, HackRF, and BladeRF. All the SDRs, as mentioned above, are capable of sample rates of over 4 MSPS, which would be enough to decode the Mode S uplink data block. Therefore dupm1090-fa could work as a base for implementing Mode S uplink data block decoding for \dumptenthirty{} in the future. 
According to the 2019 OpenSky report OpenSky network consists of 2000 receivers covering the whole of Europe and some other parts of the world ~\cite{schafer2019opensky}. Receivers can capture all types of downlink ADS-B messages in large quantities for later research usage. Uplink data is not collected because it is on a different frequency than downlink messages, and receivers can usually lack line of sight to ground stations sending interrogations~\cite{schafer2016opensky}. The amount of data would also increase significantly. However, capturing uplink messages could help to complete the whole picture of airline communications research. 

Cybersecurity and digital privacy aspects of Mode S and ADS-B are increasingly important topics, 
and recent works extensively addressed them concerning 1090MHz links 
and implementations~\cite{costin2012ghost,strohmeier2014realities,khandker2021cybersecurity,khandker2022cybersecurity-adsb}. 
As 1030MHz Mode S and ADS-B are actively used by radar systems, several works studied the security issues in radar systems~\cite{cohen2019security,nassi2021sok}, while some proposed protective mechanisms for data manipulation in radar systems~\cite{cohen2022radarnomaly}. 
However, the cybersecurity studies of the 1030MHz Mode S and ADS-B communications were not performed; therefore, we hope \dumptenthirty{} can be instrumental for this, and we leave our study of this topic as future work.


\section{Conclusion}
\label{sec:concl}

In this paper, we presented the development and evaluation of \dumptenthirty{} -- the first and only (to date) cross-platform open-source implementation decoding 1030MHz Mode A/C/S interrogations. 
In our repeated experiments, \dumptenthirty{} achieves a high detection accuracy of 1030MHz interrogation signals based on lab evaluation using synthetically-generated interrogation signals. 
We also proposed and discussed in detail a handful of practical use cases where \dumptenthirty{} can find immediate application and implementation, both in research and industrial settings.
There are several Immediate future work direction. 
The first direction is to implement pure Mode S demodulation and decoding capability assuming more capable SDR receiver frontends (e.g., HackRF). 
The second direction is to perform a more comprehensive evaluations (both longitudinal and latitudinal studies), both related to 1030MHz communications by real aircraft and transponder devices, as well as the performance evaluation of \dumptenthirty{} in the real-world. 

To further encourage validation and improvements of our work (e.g., adding real-time decoding support for Mode S P6 data block), we release \dumptenthirty{} and its artefacts as open-source. 
Moreover, we hope the ideas in this paper and the open-source release will encourage both researchers 
and practitioners across fields to apply \dumptenthirty{} in both research and industrial settings 
(including aerospace, data collection, digital privacy, and cybersecurity). 



\section*{Acknowledgment}
\label{sec:ack}

Significant parts of this research were supported by a grant from the 
\emph{Decision of the Research Dean on research funding within the Faculty (07.04.2021)} 
of the Faculty of Information Technology of the University of Jyv\"{a}skyl\"{a} 
(The authors thank Dr. Andrei Costin for facilitating and managing the grant). 

The authors acknowledge the grants of computer capacity from the 
Finnish Grid and Cloud Infrastructure (persistent identifier 
\emph{urn:nbn:fi:research-infras-2016072533}).

Hannu Turtiainen also thanks 
%
the Finnish Cultural Foundation / Suomen Kulttuurirahasto (https://skr.fi/en) 
for supporting his Ph.D. dissertation work and research (under grant decision no.00221059) 
%
and the Faculty of Information Technology of the University of Jyv\"{a}skyl\"{a} (JYU), 
in particular, Prof. Timo H\"{a}m\"{a}l\"{a}inen, for partly supporting and supervising his Ph.D. work at JYU in 2021--2022.

The authors acknowledge fair use of graphical materials courtesy of \url{www.flaticon.com} (authors: Uniconlabs, Slidicon, Ghozi Muhtarom, Freepik, mynamepong, Sweetline Graphic, Good Ware, Bombasticon Studio, and DinosoftLabs). 

Last but not least, the authors would like to thank the anonymous reviewers for their valuable feedback 
and insightful comments that meaningfully improved the quality of the final version of this paper. 


\bibliographystyle{IEEEtran}

\bibliography{FRUCTexample}

\begin{thebibliography}{10}
\providecommand{\url}[1]{#1}
\csname url@samestyle\endcsname
\providecommand{\newblock}{\relax}
\providecommand{\bibinfo}[2]{#2}
\providecommand{\BIBentrySTDinterwordspacing}{\spaceskip=0pt\relax}
\providecommand{\BIBentryALTinterwordstretchfactor}{4}
\providecommand{\BIBentryALTinterwordspacing}{\spaceskip=\fontdimen2\font plus
\BIBentryALTinterwordstretchfactor\fontdimen3\font minus
  \fontdimen4\font\relax}
\providecommand{\BIBforeignlanguage}[2]{{%
\expandafter\ifx\csname l@#1\endcsname\relax
\typeout{** WARNING: IEEEtran.bst: No hyphenation pattern has been}%
\typeout{** loaded for the language `#1'. Using the pattern for}%
\typeout{** the default language instead.}%
\else
\language=\csname l@#1\endcsname
\fi
#2}}
\providecommand{\BIBdecl}{\relax}
\BIBdecl

\bibitem{costin2012ghost}
A.~Costin and A.~Francillon, ``{Ghost in the Air (Traffic): On insecurity of
  ADS-B protocol and practical attacks on ADS-B devices},'' \emph{Black Hat
  USA}, 2012.

\bibitem{khandker2021cybersecurity}
S.~Khandker, H.~Turtiainen, A.~Costin, and T.~Hamalainen, ``{Cybersecurity
  attacks on software logic and error handling within ADS-B implementations:
  Systematic testing of resilience and countermeasures},'' \emph{IEEE
  Transactions on Aerospace and Electronic Systems}, 2021.

\bibitem{khandker2022cybersecurity-adsb}
------, ``{On the (In)Security of 1090ES and UAT978 Mobile Cockpit Information
  Systems -- An Attacker Perspective on the Availability of ADS-B Safety- and
  Mission-Critical Systems},'' \emph{IEEE Access}, 2022.

\bibitem{juvonen2022log4j}
A.~Juvonen, A.~Costin, H.~Turtiainen, and T.~Hamalainen, ``{On Apache Log4j2
  exploitation in aeronautical, maritime, and aerospace communication},''
  \emph{IEEE Access}, 2022.

\bibitem{sanfilippo2013dump1090}
S.~Sanfilippo, ``{Dump1090 is a simple Mode S decoder for RTLSDR devices},''
  2013.

\bibitem{jowettgithub}
O.~Jowett, ``Github-mutability/dump1090.''

\bibitem{rtl1030}
JetVision.de, ``Rtl1030 -- variable frequency data analyzer and 1030 mhz
  decoder, includes sisex (signals-in-space explorer),''
  \url{https://rtl1090.com/}.

\bibitem{strohmeier2015opensky}
M.~Strohmeier, I.~Martinovic, M.~Fuchs, M.~Sch{\"a}fer, and V.~Lenders,
  ``{OpenSky: A swiss army knife for air traffic security research},'' in
  \emph{AIAA 34th Digital Avionics Systems Conference}.\hskip 1em plus 0.5em
  minus 0.4em\relax IEEE, 2015.

\bibitem{sun20211090}
J.~Sun, ``{The 1090 Megahertz Riddle: A Guide to Decoding Mode S and ADS-B
  Signals},'' \emph{TU Delft OPEN Publishing}, 2021.

\bibitem{srvsop-icao}
{Sistema Regional de Cooperación para la Vigilancia de la Seguridad
  Operacional (SRVSOP)}, ``{Compliance Checklist (CC) / Electronic Filing of
  Differences (EFOD)},''
  \url{https://www.srvsop.aero/site/wp-content/uploads/2021/05/EFOD_LAR210_AN10Vol.4.pdf},
  2021, accessed: 2022-02-20.

\bibitem{traficom2021}
\emph{Radio Frequency Regulation 4}, TRAFICOM/185774/03.04.05.00/2021, Dec
  2021.

\bibitem{strohmeier2022building}
M.~Strohmeier, L.~Granger, G.~Tresoldi, and V.~Lenders, ``{Building an Avionics
  Laboratory for Cybersecurity Testing},'' in \emph{15th Workshop on Cyber
  Security Experimentation and Test (CSET) (To Appear)}, 2022.

\bibitem{khandker2022cybersecurity-ais}
S.~Khandker, H.~Turtiainen, A.~Costin, and T.~H{\"a}m{\"a}l{\"a}inen,
  ``Cybersecurity attacks on software logic and error handling within ais
  implementations: A systematic testing of resilience,'' \emph{IEEE Access},
  vol.~10, 2022.

\bibitem{hackrf}
{Great Scott Gadgets}, ``{HackRF Product Line},''
  \url{https://greatscottgadgets.com/hackrf/}.

\bibitem{grc}
{GNU Radio Project}, ``{GNU Radio Companion},''
  \url{https://wiki.gnuradio.org/index.php/GNURadioCompanion}.

\bibitem{Bredemeyer}
J.~Bredemeyer, ``{Flight Inspection of The Secondary Surveillance Radar
  Signal-In-Space},''
  \url{http://www.icasc.co/sites/faa/uploads/documents/resources/12th_int_flight_inspection_symposium/Flight_Inspection_of_the_Secondary_Surveillance_Radar_Signal-In-Space.pdf},
  2002, accessed: 2022-01-02.

\bibitem{Mateu}
J.~Mateu and J.~Berenguer, ``{Secondary Surveillance RADAR (SSR), Air Traffic
  Control Radar Beacon System (ATCRBS)},''
  \url{https://upcommons.upc.edu/bitstream/handle/2117/340881/SSR.pdf},
  accessed: 2022-02-10.

\bibitem{9092398}
J.~Kaderka and T.~Urbanec, ``Time and sample rate synchronization of rtl-sdr
  using a gps receiver,'' in \emph{2020 30th International Conference
  Radioelektronika (RADIOELEKTRONIKA)}, 2020, pp. 1--4.

\bibitem{8356821}
D.~Ball, N.~Naik, and P.~Jenkins, ``Lightweight and cost-effective spectrum
  analyser based on software defined radio and raspberry pi,'' in \emph{2017
  European Modelling Symposium (EMS)}, 2017, pp. 260--266.

\bibitem{danymol2013real}
R.~Danymol, T.~Ajitha, and R.~Gandhiraj, ``Real-time communication system
  design using rtl-sdr and raspberry pi,'' in \emph{2013 International
  Conference on Advanced Computing and Communication Systems}.\hskip 1em plus
  0.5em minus 0.4em\relax IEEE, 2013, pp. 1--5.

\bibitem{sruthi2013low}
M.~Sruthi, M.~Abirami, A.~Manikkoth, R.~Gandhiraj, and K.~Soman, ``Low cost
  digital transceiver design for software defined radio using rtl-sdr,'' in
  \emph{2013 international mutli-conference on automation, computing,
  communication, control and compressed sensing (iMac4s)}.\hskip 1em plus 0.5em
  minus 0.4em\relax IEEE, 2013, pp. 852--855.

\bibitem{aguilar2020comparative}
R.~Aguilar-Gonzalez, A.~Prieto-Guerrero, V.~Ramos, E.~Santos-Luna, and
  M.~Lopez-Benitez, ``A comparative study of rtl-sdr dongles from the
  perspective of the final consumer,'' in \emph{2020 IEEE International
  Conference on Consumer Electronics (ICCE)}.\hskip 1em plus 0.5em minus
  0.4em\relax IEEE, 2020, pp. 1--5.

\bibitem{flightradar}
A.~Flightradar24, ``{Flight Radar 24 Live Air Traffic},'' \emph{web page at:
  http://www. flightradar24. com}, 2013.

\bibitem{schafer2016opensky}
M.~Sch{\"a}fer, M.~Strohmeier, M.~Smith, M.~Fuchs, R.~Pinheiro, V.~Lenders, and
  I.~Martinovic, ``{OpenSky report 2016: Facts and figures on SSR mode S and
  ADS-B usage},'' in \emph{AIAA 35th Digital Avionics Systems
  Conference}.\hskip 1em plus 0.5em minus 0.4em\relax IEEE, 2016.

\bibitem{schafer2019opensky}
M.~Sch{\"a}fer, X.~Olive, M.~Strohmeier, M.~Smith, I.~Martinovic, and
  V.~Lenders, ``{OpenSky report 2019: Analysing TCAS in the real world using
  big data},'' in \emph{AIAA 38th Digital Avionics Systems Conference}.\hskip
  1em plus 0.5em minus 0.4em\relax IEEE, 2019.

\bibitem{harman1996beacon}
W.~H. Harman and M.~J. Brennan, ``{Beacon Radar and TCAS Interrogation Rates:
  Airborne Measurements in the 1030 MHz Band},'' MASSACHUSETTS INST OF TECH
  LEXINGTON LINCOLN LAB, Tech. Rep., 1996.

\bibitem{panken2012measurements}
A.~Panken, W.~Harman, C.~Rose, A.~Drumm, B.~Chludzinski, T.~Elder, and
  T.~Murphy, ``{Measurements of the 1030 and 1090 MHz environments at JFK
  international airport},'' \emph{Project Report ATC-390, Lincoln Laboratory
  Massachusetts Institute of Technology}, 2012.

\bibitem{cohen2019security}
S.~Cohen, T.~Gluck, Y.~Elovici, and A.~Shabtai, ``Security analysis of radar
  systems,'' in \emph{Proceedings of the ACM Workshop on Cyber-Physical Systems
  Security \& Privacy}, 2019, pp. 3--14.

\bibitem{cohen2implementation}
B.~Cohen and A.~Smith, ``{Implementation of a low-cost SSR/ADS-B aircraft
  receiver decoder (SY-100)},'' in \emph{AIAA 17th Digital Avionics Systems
  Conference}, vol.~2.\hskip 1em plus 0.5em minus 0.4em\relax IEEE, 2002.

\bibitem{strohmeier2014realities}
M.~Strohmeier, M.~Sch{\"a}fer, V.~Lenders, and I.~Martinovic, ``{Realities and
  challenges of nextgen air traffic management: the case of ADS-B},''
  \emph{IEEE Communications Magazine}, vol.~52, no.~5, pp. 111--118, 2014.

\bibitem{nassi2021sok}
B.~Nassi, R.~Bitton, R.~Masuoka, A.~Shabtai, and Y.~Elovici, ``{SoK: Security
  and privacy in the age of commercial drones},'' in \emph{2021 IEEE Symposium
  on Security and Privacy (SP)}.\hskip 1em plus 0.5em minus 0.4em\relax IEEE,
  2021, pp. 1434--1451.

\bibitem{cohen2022radarnomaly}
S.~Cohen, E.~Levy, A.~Shaked, T.~Cohen, Y.~Elovici, and A.~Shabtai,
  ``{RadArnomaly: Protecting Radar Systems from Data Manipulation Attacks},''
  \emph{Sensors}, vol.~22, no.~11, p. 4259, 2022.

\end{thebibliography}






\end{document}